\documentclass{jfm}
\usepackage{graphicx}
\usepackage{mathrsfs}
\usepackage[citation-order]{amsrefs}

\usepackage{amsmath, amssymb, graphics}
\usepackage{latexsym, amsfonts, amssymb, txfonts, pxfonts, wasysym}

\RequirePackage{amsmath} \RequirePackage{amssymb}

\ifCUPmtlplainloaded \else
  \checkfont{eurm10}
  \iffontfound
    \IfFileExists{upmath.sty}
      {\typeout{^^JFound AMS Euler Roman fonts on the system,
                   using the 'upmath' package.^^J}%
       \usepackage{upmath}}
      {\typeout{^^JFound AMS Euler Roman fonts on the system, but you
                   don't seem to have the}%
       \typeout{'upmath' package installed. JFM.cls can take advantage
                 of these fonts,^^Jif you use 'upmath' package.^^J}%
      }
  \else
  \fi
\fi

\ifCUPmtlplainloaded \else
  \checkfont{msam10}
  \iffontfound
    \IfFileExists{amssymb.sty}
      {\typeout{^^JFound AMS Symbol fonts on the system, using the
                'amssymb' package.^^J}%
       \usepackage{amssymb}%
         \let\leq=\leqslant
         
      }{}
  \fi
\fi

\ifCUPmtlplainloaded \else
  \IfFileExists{amsbsy.sty}
    {\typeout{^^JFound the 'amsbsy' package on the system, using it.^^J}%
     \usepackage{amsbsy}}
    {}
\fi

\def\Real{\mbox{Re}}      % cf plain TeX's \Re and Reynolds number
\def\i{\mbox{i}}          % cf plain TeX's \i complex number
\def\hexnumber#1{\ifcase#1 0\or1\or2\or3\or4\or5\or6\or7\or8\or9\or
 A\or B\or C\or D\or E\or F\fi }
\makeatletter \ifx\CUP@mtlplain@loaded\undefined
\else
\fi \makeatother
\newsavebox{\thalfbox}
\sbox{\thalfbox}{$\textstyle\frac{1}{2}$}

\newsavebox{\shalfbox}
\sbox{\shalfbox}{$\scriptstyle\frac{1}{2}$}

\newsavebox{\squartbox}
\sbox{\squartbox}{$\frac{1}{4}$} %RM removed scriptstyle

\newsavebox{\etbox}
\sbox{\etbox}{\boldmath$\eta$}
\newsavebox{\astrutbox}
\sbox{\astrutbox}{\rule[-5pt]{0pt}{20pt}}

\mathchardef\varLambda="0103
\makeatletter \ifx\CUP@mtlplain@loaded\undefined
\else
\fi \makeatother
\makeatletter \ifx\CUP@mtlplain@loaded\undefined
  \font\tenbms=cmbsy10
  \font\sevenbms=cmbsy10 at 7pt
  \font\fivebms=cmbsy10 at 5pt
  \newfam\bmsfam
  \textfont\bmsfam=\tenbms
  \scriptfont\bmsfam=\sevenbms
  \scriptscriptfont\bmsfam=\fivebms
  
  \edef\bsy@{\hexnumber\bmsfam}
  \mathchardef\bnabla="0\bsy@72
\fi \makeatother

\font\msym=msym10
\newfam\msfam  \textfont\msfam=\msym

\newcommand{\beqa}{\begin{eqnarray}}
\newcommand{\eeqa}{\end{eqnarray}}
\newcommand{\no}{\nonumber}

\newcommand{\df}[2]{\displaystyle\frac{#1}{#2}}
\newcommand{\tf}[2]{\textstyle\frac{#1}{#2}}

\newcommand{\Int}[2]{\displaystyle\int_{#1}^{#2}}

\newcommand{\PDD}[2]{\df{\partial #1}{\partial #2}}

\newcommand{\PDDT}[2]{\df{\partial^{2} #1}{\partial #2^{2}}}

\newcommand{\PTT}[1]{\df{\partial^{2} }{\partial #1^{2}}}

\newcommand{\os}[1]{\overline{#1}}

\newcommand{\be}{\begin{eqnarray}}
\newcommand{\en}{\end{eqnarray}}
\newcommand{\Half}{\mbox{\tiny $\tf{1}{2}$}}
\newcommand{\half}{\mbox{\footnotesize $\tf{1}{2}$}}

\title[Growth of Stokes Waves by Wind on a Viscous Liquid]
{Growth of Stokes Waves Induced by Wind on a Viscous Liquid of Infinite Depth}

\author[S. G. Sajjadi]%
{S\ls H\ls A\ls H\ls R\ls D\ls A\ls D\ns G.\ns
S\ls A\ls J\ls J\ls A\ls D\ls I}

\affiliation{
Department of Mathematics, ERAU,\ Florida, USA,\\
and Trinity College, Cambridge, UK.}
\pubyear{2003} \volume{268} \pagerange{1--30}
\date{April 2016}
\begin{document}
\maketitle

\begin{abstract}
\footnotesize{The original investigation of Lamb (1932, \S 349) for the effect of viscosity on monochromatic surface waves is
extended to account for second-order Stokes surface waves on deep water in the presence of surface tension. This extension is used to evaluate interfacial impedance
for Stokes waves under the assumption that the waves are growing and hence  the surface waves are unsteady. 
Thus, the previous investigation of Sajjadi {\it et al}. (2014) is further explored in that (i) the surface wave is unsteady and nonlinear, and (ii) the effect of the water viscosity, which affects surface stresses, is taken into account. 
The determination of energy-transfer parameter, from wind to waves, are calculated through a turbulence closure model but it is shown the contribution due to turbulent shear flow is some 20\% lower than that obtained previously.  
A derivation leading to an expression for the closed streamlines (Kelvin cat-eyes), which arise in the vicinity of the critical height, is found for unsteady surface waves. From this expression it is deduced that as waves grow or decay, the cats-eye are no longer symmetrical.  
Also investigated is the energy transfer from wind to short Stokes waves through the viscous Reynolds stresses in the immediate neighborhood of the water surface.  It is shown that the resonance between the Tollmien-Schlichting waves for a given turbulent wind velocity profile and the free-surface Stokes waves give rise to an additional contribution to the growth of nonlinear surface waves.}
\end{abstract}

\section{Introduction}

The energy exchange from wind to waves crucially depends on
accurate determination of stresses on the water surface. The
energy-transfer parameter (as is commonly known in the literature)
is determined from the complex part of the interfacial impedance
(as is termed by Miles), see (\ref{B5}) below. John Miles made
several analytical attempts to improve upon the energy-transfer
parameter, beginning with his pioneering work in 1957 and his
final contribution in 1996. In all his contributions he assumed
the initial surface is composed of a monochromatic surface wave of
small steepness. Miles (2004) remarked "... it will be interesting
to see the extension of my 1996 contribution to Stokes waves, and
its comparison with some numerical studies."  Although it was not
explicitly mentioned by Miles, we assume he was referring to
numerical contributions, for example, by Al-Zanaidi \& Hui (1984)
and Mastenbroek {\em et al.} (1996). However, he had recognized a
major obstacle for this task, and further commented "... the
proper determination of the interfacial impedance for an {\em a
priori} assumed nonlinear waves is by no means
straightforward...". In this note we offer a way to resolve this
anomaly.

In a recent study by Sajjadi, Hunt, and Druillion (2014), it was shown that the growth rate, $kc_i$, where $k$ is the wave number and $c_i$ is the wave complex part of the wave phase speed, for growing waves critically depend on the energy-transfer parameter $\beta$.  Moreover, Sajjadi and Hunt (2003) (SHD therein) have suggested that wave steepness (for nonlinear waves, such as those observed in the sea) are also a contributing factor for the momentum transfer from wind to surface waves, see also Sajjadi (2015). 
Thus, one goal of the present study is the accurate determination of $\beta$, and this requires the calculation of the complex amplitude of the wave-induced pressure at the surface. 

To achieve our objective we must extend the effect of viscosity on monochromatic waves (Lamb 1932, \S 349) to a nonlinear surface wave, here we shall consider the simplest case, i.e. that of the Stokes wave.  Thus, we shall adopt the bicrohomic assumption for the mean motion which admits the representation of the form (see section 4)
$$
(\sigma, \chi, \tau)=\Real\left\{({\mathscr P}_1, \Xi_1, {\mathscr T}_1)e^{ik(x-ct)}
+ka({\mathscr P}_2, \Xi_2, {\mathscr T}_2)e^{2ik(x-ct)}\right\}
$$
With ${\mathscr P}_1, \cdots$, representing complex amplitude of stresses, and where $c=c_r+ic_i$ ($c_r$ is the wave speed) and $ka$ is the wave steepness.

For an unsteady monochromatic surface wave SHD showed that the total energy transfer comprises the sum of two components $\beta_c+\beta_T$, 
\be 
\beta_c=-\pi(U_c''/kU_c')(\os{{\mathscr W}_c^2}/U_1^2 h^2_{0x}),\qquad (\,\,\,)'\equiv\df{d}{dz}\label{1.1}
\en 
is the contribution associated with the singularity at the critical layer Miles (1957), but due to the unsteadiness of the surface wave $\beta_c$ is additionally a function of $c_i$ (see SHD).  In (1.1), the overbar signifies an average over $x, U_1$ is the kinematic friction velocity, ${\cal W}$ is the wave-induced vertical velocity, and the subscript $c$ denotes evaluation at the critical point $z=z_c$, where $U(z_c)=c_r$.  The second component, $\beta_T$ is the rate of energy transfer to the surface due to the turbulent shear flow blowing over it. 

Thus, we determine the energy-transfer parameter through calculating the pressure $
{\mathscr P}_\ell$, and the shear stress ${\mathscr T}_\ell$, $(\ell=1, 2)$ at the surface.  As mentioned above, this requires generalization of the interfaced impedance by extending the monochromatic viscous theory of water waves (Lamb 1932 \S 349) to account for the effect of viscosity for Stokes waves.  Then, the extended Lamb’s solution to Stokes wave in a viscous liquid, with prescribed stresses at the surface can be adopted to evaluate expressions for $\beta_c$ and $\beta_T$ in the limit as $s\equiv \rho_a/\rho_w\downarrow 0$, as outlined in section 3 below.

Hence, following the procedure adopted by SHD, for unsteady monochromatic waves, which led through evaluation of ${\mathscr P}_0$  and ${\mathscr T}_0$, to the following expressions
\be 
& &\beta_c=\pi\xi_c^3L_0^4\left[1+(4-\tf{1}{3}\pi^2+10\hat{c}_i^2)\Lambda^2+O(\Lambda^2)\right]\label{1.2}\\
& &\beta_T=5\kappa^2L_0+O(\Lambda)\label{1.3}
\en 
is obtained for an unsteady Stokes wave in sections 4 and 5.  In equations (1.2) and (1.3) $\xi_c=kz_c$, $L_0=\gamma-\ln(2\xi_c)$ and $\gamma=0.5772$ is Euler’s constant. 

In section 6, we derive an expression for closed loop streamlines, namely Kelvin’s cat-eye, and the significance of which is explored and explained. Finally, the results and discussion is given in section 7.

\section{Interfacial impedance}

Miles-Sajjadi (Miles 1996, Sajjadi 1998, hereafter M96 and S98 respectively) theory of surface wave generation
considers the role of wave-induced Reynolds stresses in the transfer of energy from a turbulent shear flow to
gravity waves on deep water.  In their theories the Reynolds-averaged equations for turbulent flow over a deep-water
 sinusoidal gravity wave, $z=a\cos kx\equiv h_0(x)$ (M96) and fully nonlinear surface gravity wave
  $z=\sum^{\infty}_{n=1}a_n\cos k_n x$ (S98), are formulated.  Their formulations uses the wave-following
  coordinates $\xi, \eta$, where  $x=\xi, z=\eta+h(\xi,\eta), h(\xi,0)=h_0(\xi)$ and $h$ is exponentially
   small for $k\eta\gg 1$.  The turbulent Reynolds stress equations are closed by a viscoelastic constituent
   equation-a mixing-length model with relaxation (M96) and by the rapid-distortion theory (S98).  Both derive
   their evolution equation on the assumption that: (i) the basic velocity profile is logarithmic in $\eta+z_0$,
    where $z_0$ is a roughness length; (ii) the lateral transport of turbulent energy in the perturbed flow
    is negligible; and (iii) the dissipation length is proportional to $\eta+z_0$.  In both theories an
    inhomogeneous counterpart of the Orr-Sommerfeld equation is derived for the complex amplitude of
     the perturbation streamfunction and then used to construct a quadratic functional for the energy
      transfer to the wave.  A corresponding Galerkin approximation that is based on independent variational
      approximation for outer (quasi-laminar) and inner (shear-stress) domains yields the interfacial
      impedance $\alpha+\i\beta$ (defined by Miles 1957) in the limit $s\equiv\rho_{\rm a}/\rho_{\rm w}\downarrow 0$.
       The calculation of the interfacial impedance requires the solution of the linearized equations of motion of
        water bounded above by a monochromatic surface wave (Lamb 1932, \S 349).

However, for Stokes surface wave $z=a\cos kx+\tf{1}{2} ka^2\cos 2kx$ the extension of Lamb's solution is not
 immediately obvious (and Lamb, as well as other researchers to date, did not address this problem). However,
 in the case of a shear flow over a sinusoidal wave (for the application to air-sea interactions), we may
 consider the solution of the linearized Navier-Stokes equations in the semi-infinite body of water bounded above by the surface wave
\be
z=a{\rm e}^{{\rm i} k(x-ct)}\equiv h_0(x,t)\qquad(ka\ll 1)\label{B1}
\en
in a fixed frame of reference gives, after renaming variables, that is after letting
 $A=\i ac(1+\mathscr{C}), C=ac\mathscr{C}, n=-\i kc$, (where the letters on the left-hand
 side denote those used by Lamb), and neglecting surface tension therein; following Lamb and adopting complex dependent variables, we obtain
\be
u_{\rm w}=[k+\mathscr{C}(k-m)]ch_0\label{B2a}
\en
\be
\tau_{\rm w}=(\tau_{13})_{\rm w}=(2\nu_{\rm w} kc+{\rm i}\mathscr{C}c^2)kh_0\label{B2b}
\en
and
\be
\left(-\df{p}{\rho}+\tau_{33}\right)_{\rm w}=\left\{\df{g}{k}-c^2-2{\rm i}k\nu_{\rm w}c-\mathscr{C}[c^2-2{\rm i}(k-m)\nu_{\rm w} c]\right\}kh_0\label{B2c}
\en
for the tangential velocity, tangential stress, and normal stress, respectively, at the surface.
 The subscript $w$ refers to water, $m\equiv [k^2-\i(kc/\nu_{\rm w})]^{\Half}$, and
$s=\rho_a/\rho_w$ being the ratio of air to water density.

Invoking continuity of the perturbation velocity $u=c\partial h/\partial\eta$ and $\tau$ and
$-\sigma$, the tangential and normal stresses, eliminating $\mathscr{C}$, and letting
 $k\nu_{\rm w}/c\downarrow 0$, we obtain the interfacial conditions\footnote{We emphasize that, $u$ is perturbation to tangential wind velocity and $c$ is the complex phase speed induced by the presence of the wind.}
\be
u-{\rm i}(kc\nu_{\rm w})^{-\Half}s\tau=kch_0\label{B3a}
\en
\be
s(\sigma+{\rm i}\tau)=(c^2-c^2_{\rm w})kh_0\label{B3b}
\en
where
\be
c_{\rm w}\equiv(g/k)^{\Half}-2{\rm i}k\nu_{\rm w}\qquad(|k\nu_{\rm w}/c|\ll 1)\label{B4}
\en
is the complex phase speed in the absence of the air comprehends (through its imaginary part, which
 may be replaced by an empirical equivalent) the dissipation in water.  The ratio of the second term
 to the first term on the left-hand side of (\ref{B3a}) is typically smaller than $10^{-2}$;
 accordingly (\ref{B3a}) may be approximated by $u=kch_0$.  However, we note that (\ref{B3a})
 does not reduce to $c=kch_0$ in the limit of air inviscid liquid $(k\nu_{\rm w}\rightarrow 0)$.

Finally by replacing $\sigma$ and $\tau$ with their complex amplitudes $\mathscr{P}$ and $\mathscr{T}$
(defined as in Miles 1957) the interfacial impedance is obtained which may be expressed in the form
\be
\alpha+{\rm i}\beta\equiv\df{c^2-c_{\rm w}^2}{sU^2_1}=\df{({\mathscr P}+{\rm i}{\mathscr T})_0}{kaU^2_1}\label{B5}
\en
where $U_1=U_*/\kappa, U_*$ is the kinematic shear stress, $\kappa\approx 0.4$ is von K\'arm\'an's constant,
 and the suffix zero indicates evaluation on $\eta=0$, which to $O(ka)\ll 1$ is the same as evaluation at $z=0$.

Sajjadi (1998) followed Miles (1996) and calculated the
interfacial impedance for every harmonic of the fully nonlinear
surface wave.  We remark that although M96 and S98 formulations are
basically different for the turbulent flow over a surface wave,
nevertheless the form of the interfacial impedance adopted are the
same, provided Sajjadi's series, for the representation of a fully
nonlinear surface wave, is truncated after the first harmonic.
Moreover, in S98, the inclusion of surface tension leads to an
ambiguous results, even for the second-order Stokes wave, when his
series is truncated after the second harmonic.

This ambiguity can readily be seen from equation (\ref{12}) below in which
\be
(\phi, \zeta, w)=\sum_{s}(\Phi_s, \Xi_s, W_s)\no
\en
where
\be
\Phi_s=A_s{\rm e}^{k_sz},\qquad \Xi_s=\left(A_s{\rm e}^{k_sz}-{\rm i}C_s{\rm e}^{m_sz}\right),\qquad W_s=-k_s\Xi_s,\no
\en
then substituting into (\ref{12}) then for the $s$-harmonic we have
\be
\mbox{[LHS (\ref{12})]}=-\df{1}{n_s}\left\{[n_s^2+k_s(g+{\cal T}k^2)+2\nu_wk_s^2n_s]A_s-{\rm i}k_s[2\nu_wm_sn_s+g+{\cal T}
k^2]C_s\right\}\no
\en
Here distinction has to be made between $k$, the wavenumber of the entire wave, and $k_s$, the wavenumber
associated with the $s$-harmonic. One may make the seemingly obvious assumption that $k_s=sk$, however,
this will lead to a result
which appears to be wrong. Note incidently, this ambiguity can be circumvented if surface tension is neglected as in S98.

The purpose of this note is to extend Lamb's original investigation, for monochromatic waves, to Stokes waves in the presence of surface tension but under the assumption that the wave steepness $ka\ll 1$.

\section{Stokes waves on a viscous liquid}
We consider the effect of viscosity on Stokes waves on deep water whose profile is given by
\be
z=a{\rm e}^{{\rm i}k(x-ct)}+\tf{1}{2}ka^2{\rm e}^{2{\rm i}k(x-ct)}\equiv h_0+kah_1
=h(x,t),\qquad(ka\ll 1)\label{0}
\en
see figure 1.

\begin{figure}
   \begin{center}
\includegraphics[width=12cm]{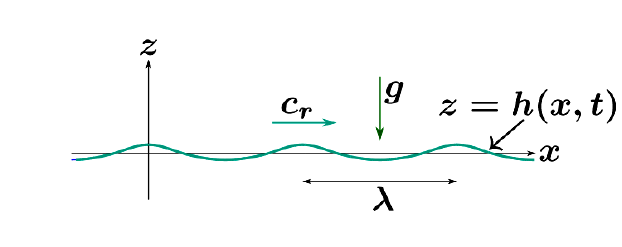}
   \end{center}
\caption{\footnotesize Schematic diagram showing a Stokes wave.}
\end{figure}

If we take the $z$-axis to be vertically upwards, and if we assume a two-dimensional motion with velocities 
${\sf u}$ and ${\sf w}$
being confined to the $x, z$-coordinates and pressure ${\sf p}$, then ignoring the inertia terms, the equations of motion
may be cast as
\be
\PDD{\sf u}{t}=-\df{1}{\rho_{\rm w}}\PDD{\sf p}{x}+\nu_{\rm w}\nabla^2{\sf u},\qquad
\PDD{\sf w}{t}=-\df{1}{\rho_{\rm w}}\PDD{\sf p}{z}-g+\nu_{\rm w}\nabla^2{\sf w},
\label{1}
\en
together with the continuity equation
\be
\PDD{\sf u}{x}+\PDD{\sf w}{z}=0\label{2}
\en
where $\rho_{\rm w}, \nu_{\rm w}$ and $g$ are the water density, the kinematic viscosity of water and acceleration due to gravity, respectively.

Equations (\ref{1}) and (\ref{2}) are satisfied by
\be
{\sf u}=-\PDD{\phi}{x}-\PDD{\psi}{z},\qquad
{\sf w}=-\PDD{\phi}{z}+\PDD{\psi}{x},
\label{3}
\en
and the linearized dynamic condition
\be
\df{\sf p}{\rho_{\rm w}}=\PDD{\phi}{t}-gz\label{4}
\en
provided
\be
\nabla_1^2\phi=0,\qquad
\PDD{\psi}{t}=\nu_{\rm w}\nabla^2_1\psi,
\label{5}
\en
where
$$\nabla^2_1\equiv\PTT{x}+\PTT{z}.$$

We consider the solutions in normal mode by assuming that they are periodic in $x$
with a prescribed wavelength $\lambda=2\pi/k$. Thus, assuming transient factors ${\rm e}^{nt}, {\rm e}^{2nt}$
and spacial factors ${\rm e}^{{\rm i}kx}, {\rm e}^{2{\rm i}kx}$, for first and second harmonics, respectively. The solution
of (\ref{5}) may therefore be expressed in the following form:
\be
\left.\begin{array}{l}
\phi=(A_1{\rm e}^{kz}+B_1{\rm e}^{-kz}){\rm e}^{{\rm i}kx+nt}+\tf{1}{4}ka(A_2{\rm e}^{2kz}+B_2{\rm e}^{-2kz}){\rm e}^{2({\rm i}kx+nt)}\\
\\
\psi=(C_1{\rm e}^{mz}+D_1{\rm e}^{-mz}){\rm e}^{{\rm i}kx+nt}+\tf{1}{4}ka(C_2{\rm e}^{2mz}+D_2{\rm e}^{-2mz}){\rm e}^{2({\rm i}kx+nt)}\\
\end{array}\right\}\label{6}
\en
with
\be
m^2=k^2+n/\nu_{\rm w}\label{7}
\en
The boundary conditions will provide equations which are sufficient to determine the nature of the
various modes, and the corresponding values of $n$.

In the case of infinite depth one of these conditions takes the form that the motion must be finite
at $z=-\infty$. Excluding for the present case where $m$ is purely imaginary, this requires that
$B_i=D_i=0$ for $i=1,2$ provided $m$ denote the root of equation (\ref{7}) with $\Real\{m\}>0$.
Hence
\be
\left.\begin{array}{l}
{\sf u}=-({\rm i}kA_1{\rm e}^{kz}+mC_1{\rm e}^{mz}){\rm e}^{{\rm i}kx+nt}-\tf{1}{2}ka({\rm i}kA_2{\rm e}^{2kz}+mC_2{\rm e}^{2mz}){\rm e}^{2({\rm i}kx+nt)}\\
\\
{\sf w}=-k(A_1{\rm e}^{kz}-{\rm i}C_1{\rm e}^{mz}){\rm e}^{{\rm i}kx+nt}-\tf{1}{2}k^2a(A_2{\rm e}^{2kz}-{\rm i}C_2{\rm e}^{2mz}){\rm e}^{2({\rm i}kx+nt)}\\
\end{array}\right\}\label{8}
\en

Since $h$ denotes the elevation at the free surface, then the linearized kinematic condition is
$\partial h/\partial t={\sf w}$. Taking the origin of $z$ at the undisturbed level, this condition gives
\be
h=-\df{k}{n}(A_1-{\rm i}C_1){\rm e}^{{\rm i}kx+nt}-\df{k^2a}{4n}(A_2-{\rm i}C_2){\rm e}^{2({\rm i}kx+nt)}.\label{9}
\en

Let $T$ be the surface tension, then the stress conditions at the free surface are given by
\be
p^{(zz)}=T\PDDT{h}{x},\qquad p^{(xz)}=0\label{10}
\en
to the first order, since we have assumed the inclination of the surface to the horizontal is sufficiently
small, so that $ka\ll 1$.

Now, if $\mu_{\rm w}$ denotes the dynamic viscosity of the water,
\be
p^{(zz)}=-{\sf p}+2\mu_{\rm w}\PDDT{\sf w}{z},\qquad p^{(xz)}=\mu_{\rm w}\left(\PDD{\sf w}{x}+\PDD{\sf u}{z}\right)\label{11}
\en
whence, by (\ref{4}) and (\ref{9}) we find, at the surface
\be
\df{p^{(zz)}}{\rho_{\rm w}}-{\cal T}\PDDT{\zeta}{x}=-\PDD{\phi}{t}+(g+{\cal T}k^2)\zeta+2\nu_{\rm w}\PDD{\sf w}{z}\label{12}
\en
where ${\cal T}=T/\rho_{\rm w}$. Next, writing $p^{(zz)}=p_1^{(zz)}+kap_2^{(zz)}$, $h=h_0+kah_1$, $\phi=\phi_1+ka\phi_2$
and ${\sf w}={\sf w}_1+ka{\sf w}_2$, then from (\ref{12}), (\ref{9}), (\ref{8}) and (\ref{6}) we have, after equating coefficients for
${\rm e}^{{\rm i}kx+nt}$ and ${\rm e}^{2({\rm i}kx+nt)}$, respectively,
\be
& &\df{p_1^{(zz)}}{\rho_{\rm w}}-{\cal T}\PDDT{h_0}{x}=-\df{1}{n}\left\{[n^2+k(g+{\cal T}k^2)+2\nu_{\rm w}nk^2]A_1
-{\rm i}k[2\nu_{\rm w}mn+g+{\cal T}k^2]C_1\right\}\no\\
& &\df{p_2^{(zz)}}{\rho_{\rm w}}-{\cal T}\PDDT{h_1}{x}=-\df{1}{n}\left\{\left[\df{n^2}{2}+\df{k}{4}(g+{\cal T}k^2)+2\nu_{\rm w}nk^2\right]A_2
-{\rm i}k\left[2\nu_{\rm w}mn+\tf{1}{4}(g+{\cal T}k^2)\right]C_2\right\}\no
\en
Similarly, writing $p^{(xz)}=p_1^{(xz)}+akp_2^{(xz)}$ and $u=u_1+aku_2$, we obtain from (\ref{11}) and (\ref{8})
\be
& &\df{p_1^{(xz)}}{\rho_{\rm w}}=-\{2{\rm i}\nu_{\rm w}k^2A_1+(n+2\nu_{\rm w}k^2)C_1\}\label{15}\\
& &\df{p_2^{(xz)}}{\rho_{\rm w}}=-\{2{\rm i}\nu_{\rm w}k^2A_2+(n+2\nu_{\rm w}k^2)C_2\}\label{16}
\en

Substituting (\ref{15}) into the first equation of (\ref{10}), and eliminating the ratio $A_1:C_1$, we see to $O(1)$ that
\be
(n+2\nu_{\rm w}k^2)^2+gk+{\cal T}k^3=4\nu_{\rm w}^2k^2m\label{17}
\en
Similarly, substitution of (\ref{16}) into the second equation of (\ref{10}), and eliminating the ratio $A_2:C_2$, yields, to $O(ka)$
\be
(n+4\nu_{\rm w}k^2)(n+2\nu_{\rm w}k^2)+gk+{\cal T}k^3=16\nu_{\rm w}^2k^2m\label{18}
\en
Eliminating $m$ between (\ref{17}) and (\ref{18}) then gives
$$n=-2\nu_{\rm w}k^2\pm\sqrt{4\nu_{\rm w}^2k^4-6\sigma^2}$$
and by virtue of the fact that $\nu_{\rm w}k\ll 1$, we obtain \be
n=-2\nu_{\rm w}k^2\pm {\rm i}\sigma^*\label{19} \en where
$\sigma^*=\sqrt{6}\sigma$ and $\sigma^2=gk+{\cal T}k^3$.

The condition $p^{(xz)}=0$ shows that
\be
\df{C_1}{A_1}=\df{C_2}{A_2}=-\df{2{\rm i}\nu_{\rm w}k^2}{n+2\nu_{\rm w}k^2}=\mp\df{2\nu_{\rm w}k^2}{\sigma^*}\label{20}
\en
which is, under the same circumstances, very small. Hence the motion is approximately irrotational,
with a velocity potential
\be
\phi=A_1{\rm e}^{-2\nu_{\rm w}k^2t+kz+{\rm i}(kx\pm\sigma^*t)}+\tf{1}{4}kaA_2{\rm e}^{-4\nu_{\rm w}k^2t+2kz+2{\rm i}(kx\pm\sigma^*t)}\label{21}
\en

If we put
$$A_1=A_2=\mp\df{{\rm i}\sigma^*a}{k}$$
the equation (\ref{9}) of the free surface becomes, on taking the real parts
\be
h=a{\rm e}^{-2\nu_{\rm w}k^2t}\cos(kx\pm\sigma^*t)+\tf{1}{4}a^2k{\rm e}^{-4\nu_{\rm w}k^2t}\cos 2(kx\pm\sigma^*t)\label{22}
\en

Since the motion is nearly (but not exactly) irrotational, there
is vorticity present whose magnitude is given by
$$\omega=\PDD{\sf w}{x}-\PDD{\sf u}{z}\equiv\nabla_1^2\psi$$
Thus, from (\ref{7}) and (\ref{19}), we have approximately
$$m=(1\pm {\rm i})b\quad{\rm where}\quad b=(\sigma^*/2\nu_{\rm w})^{1/2}$$
Hence, with the same notation as before, we find
\be
\omega=&\mp& 2\sigma^*ka{\rm e}^{-2\nu_{\rm w}k^2t+b z}\cos\{kx\pm(\sigma^*t+b z)\}\no\\
&\mp& 2\sigma^*k^2a^2{\rm e}^{-4\nu_{\rm w}k^2t+2b z}\cos 2\{kx\pm(\sigma^*t+b z)\}\label{23}
\en

From equation (\ref{23}) it can be seen that the vorticity
diminishes rapidly from the surface downwards. Moreover, since the
motion has an oscillatory character, the sign of the vorticity
which is being diffused inwards from the surface is continually
reversing, such that (paraphrasing Lamb) `beyond a stratum' of
thickness of $O(2\pi/b)$ the effect diminishes.

The above analysis gives results for the first two components of
the normal modes of the prescribed wavelength. For a fully
nonlinear Stokes wave, there are an infinitely more of these modes exist
and they correspond to pure-imaginary values of $m$, which are
less persistent in character.

It is interesting to note that, if we now, in place of (\ref{6}), assume
\be
\left.\begin{array}{l}
\phi=A_1{\rm e}^{kz}{\rm e}^{{\rm i}kx+nt}+\tf{1}{4}kaA_2{\rm e}^{2kz}{\rm e}^{2({\rm i}kx+nt)}\\
\\
\psi=(C_1\cos\ell z+D_1\sin\ell z){\rm e}^{{\rm i}kx+nt}+\tf{1}{4}ka(C_2\cos 2\ell z+D_2\sin\ell z){\rm e}^{2({\rm i}kx+nt)}\\
\end{array}\right\}\label{24}
\en
%with
%$$\ell^2=-k^2-\df{n}{\nu_{\rm w}}$$
and carrying out the previous analysis, we find to $O(1)$
\be
\left.\begin{array}{r}
(n^2+2\nu_{\rm w}k^2n+gk+{\cal T}k^3)A_1-{\rm i}(gk+{\cal T}k^3)C_1+2{\rm i}\nu_{\rm w}k\ell nD_1=0\\
\\
2{\rm i} k^2A_1+(k^2-\ell^2)C_1=0\\
\end{array}\right\}\label{25}
\en
and to $O(ka)$
\be
\left.\begin{array}{r}
(2n^2+8\nu_{\rm w}k^2n+gk+{\cal T}k^3)A_2-{\rm i}(gk+{\cal T}k^3)C_2+8{\rm i}\nu_{\rm w}k\ell nD_2=0\\
\\
2{\rm i} k^2A_2+(k^2-\ell^2)C_2=0\\
\end{array}\right\}\label{26}
\en
We note that now any value of $\ell$ is admissible in these equations for determining
the ratios $A_i:C_i:D_i, (i=1,2)$; and the corresponding value of $n$ is
$$n=-\nu_{\rm w}(k^2+\ell^2)$$

We remark the extension of the above analysis to third or higher order Stokes waves, (if at all analytically tractable) is by no means an easy task.

\section{Energy transfer to unsteady Stokes waves}

We consider a turbulent shear flow of air whose density is $\rho_a$ blowing over
an unsteady second-order Stokes wave of the form (\ref{0}) with a complex phase speed $c=c_r+ic_i$, where $c_r=\sqrt{g/k(1+k^2a^2)}$ is the wave speed and $kc_i$ is the growth ($>0$) or decay ($<0$) rate.

We shall neglect the molecular viscosity of the air by virtue of the fact that $\nu_a\ll\nu_w$ and thus the viscous forces in the airflow becomes negligible. Then, the governing Reynolds-averaged equations are given by S98
$$\partial_i\langle u_i\rangle=0,\qquad\mathscr{D}\langle u_i\rangle=-\partial_i\langle p/\rho_a\rangle
-\partial_j\langle u_i'u_j'\rangle,$$
where
$$\mathscr{D}=\partial_t+\langle u_j\rangle\partial_j.$$
Hence, the horizontal and vertical momentum equations may be expressed, respectively, as
\be 
\mathscr{D}\langle u\rangle=-\sigma_x+\chi_x+\tau_z\label{4.1}
\en 
and
\be 
\mathscr{D}\langle w\rangle=-\sigma_z+\tau_x\label{4.2}
\en 
where
$$\sigma\equiv\langle p/\rho_a+w^{\prime 2}\rangle,\qquad\chi\equiv\langle u^{\prime 2}-w^{\prime 2}\rangle,\qquad\tau\equiv -\langle u'w'\rangle.$$
Here $\sigma$ and $\tau$ are the mean normal and shear stresses.

Following Townsend's scaling argument (Townsend 1972), we may further neglect the components $\chi_x$ and $\tau_x$ without affecting the solution significantly. Accordingly, equations (\ref{4.1}) and (\ref{4.2}) reduce to
$$\mathscr{D}\langle u\rangle=-\sigma_x+\tau_z\qquad\mathscr{D}\langle w\rangle=-\sigma_z.$$
The continuity of the air-water at the surface requires
$$u=c\PDD{h}{\eta},\qquad s\tau=\tau_w\quad{\rm and}\quad s\sigma=\sigma_w$$
where $s=\rho_a/\rho_w\ll 1$. Note we have used the same transformations given in section 2. However, we note that if the wave steepness $ka\ll 1$ the by virtue of which $z\approx\eta+O(ka)$.

Using the expression for the horizontal velocity, given by the first of equations (\ref{8}), in the curvilinear coordinates, namely
\be 
{\sf u}=-({\rm i}kA_1{\rm e}^{k\eta}+mC_1{\rm e}^{m\eta}){\rm e}^{{\rm i}k\xi+nt}-\tf{1}{2}ka({\rm i}kA_2{\rm e}^{2k\eta}+mC_2{\rm e}^{2m\eta}){\rm e}^{2({\rm i}k\xi+nt)},\no 
\en 
and using the following transformation:
$$(A_1, A_2, C_1, C_2, n)=[iac(1+{\mathscr C}_1), 2iac(1+{\mathscr C}_2), ac{\mathscr C}_1,
2ac{\mathscr C}_2, -ikc]$$
we obtain
$$u_w=c[k+{\mathscr C}_1(k-m)]h_0+2kac[k+{\mathscr C}_2(k-m)]h_1,$$
where the subscript $w$ refers to water.

Neglecting the surface tension, for Stokes waves the total mean normal stress at the surface is (see section 3) 
$$\sigma_w\equiv p^{(\eta\eta)}=p_1^{(\eta\eta)}+kap^{(\eta\eta)}\equiv\left(-\df{{\sf p}}{\rho}+p_{33}\right)_w.$$
Thus, referring to the previous section, we have
\be 
& &p_1^{(\eta\eta)}=ka\left\{\df{g}{k}-c^2-2ik\nu_wc-{\mathscr C}_1[c^2-2i(k-m)\nu_wc]\right\}\no\\
& &p_2^{(\eta\eta)}=\tf{1}{2}ka\left\{\df{g}{k}-2c^2-8ik\nu_wc-2{\mathscr C}_2[c^2-4i(k-m)\nu_wc]\right\}\no 
\en 
and similarly,
$$\tau_w=p^{(\xi\eta)}=(2\nu_wk+i{\mathscr C}_1c)ckh_0+2ka(2\nu_wk+i{\mathscr C}_2c)ckh_1.$$

We next express the mean surface stresses in the bichromatic perturbation form
$$(\sigma_w, \tau_w)=({\mathscr P}_1h_0+ka{\mathscr P}_2h_1, {\mathscr T}_1h_0+ka{\mathscr T}_2h_1)$$
where ${\mathscr P}_i$ and ${\mathscr T}_i, (i=1,2)$ represent the complex amplitudes of the normal and shear stresses, respectively. Invoking continuity of the perturbation velocity ($u=c\partial h/\partial\eta$), eliminating ${\mathscr C}_1$ and letting $k\nu_w/c\downarrow 0$ (as $m\rightarrow k$), we obtain the interfacial conditions:
$$u_1-is(kc\nu_w)^{-\half}{\mathscr T}_1h_0+\tf{1}{2}ka\left[2u_2-is(kc\nu_w)^{-\half}{\mathscr T}_2h_1\right]$$
\be
s[({\mathscr P}_1+i{\mathscr T}_1)h_0+ka({\mathscr P}_2+i{\mathscr T}_2)h_1]=
(c^2-c^2_w)kh_0+ka(2c^2-c_w^2)kh_1\label{4.3}
\en 
where $c_w$ is given by (\ref{B4}). Hence from (\ref{4.3}) we see that
\be 
\df{c^2(1+2ka)-c_w^2(1+ka)}{sU_1^2}&=&\df{({\mathscr P}_1+i{\mathscr T}_1)+ka({\mathscr P}_2+i{\mathscr T}_2)}{kaU_1^2}\no\\
&=&(\alpha_1+i\beta_1)+ka(\alpha_2+i\beta_2).\label{4.4}
\en 

\section{Determination of energy-transfer parameter}

The energy-transfer parameter, $\beta$ defined as in (\ref{4.4}), requires the calculation
of the complex amplitudes of the wave-induced pressure and shear stress at the surface of
Stokes waves. We may obtain these from the solution of the Orr-Sommerfeld equations (for details
see S98 or SHD),
$$[\nu_e(\Phi_1''+U''{\mathscr H}''_1)]''={\rm i}k[{\mathscr U}(\Phi_1''-k^2\Phi_1)-U''\Phi_1]$$
$$[\nu_e(\Phi_2''+U''{\mathscr H}''_2)]''=2{\rm i}k[{\mathscr U}(\Phi_2''-4k^2\Phi_2)-U''\Phi_2]$$
where $(^{\prime}\equiv d/d\eta)$, subject to boundary conditions at $\eta=0$ and $\eta=\infty$.
In the above equations $\nu_e$ is the eddy viscosity, $\mathscr{U}\equiv U-c_r$, and $\mathscr{H}(\eta)$ and $\Phi(\eta)$ are, respectively, the complex amplitudes of $h$ and the perturbation stream function $\phi$, given by 
\be 
\Psi=\int_0^\eta\mathscr{U}(\eta)\,d\eta+\mathscr{U}h(\xi,\eta)+\phi(\xi,\eta)\label{5.1a}
\en 
with understanding that $h=h_1+kah_2$, etc.

Alternatively we can follow generalization of M96 for Stokes waves and evaluate $\beta$ by taking the real part of the quadratic functional
\be
\beta=(kaU_1)^{-2}\int^\infty_0\{\nu_e[{\mathscr U}{\mathscr
Z}_1''^2+2U'{\mathscr Z}_1'{\mathscr Z}_1''+U''({\mathscr Z}_1-{\mathscr
H}_1){\mathscr Z}_1'']+
\no\\
+ik{\mathscr U}^2({\mathscr Z}_1'^2+k^2{\mathscr Z}_1^2)\}d\eta\no\\
+ka(2kaU_1)^{-2}\int^\infty_0\{\nu_e[{\mathscr U}{\mathscr
Z}_2''^2+2U'{\mathscr Z}_2'{\mathscr Z}_2''+U''({\mathscr Z}_2-{\mathscr
H}_2){\mathscr Z}_2'']
\no\\
+2ik{\mathscr U}^2({\mathscr Z}_2'^2+4k^2{\mathscr Z}_2^2)\}d\eta,\label{5.1}
\en
which provides a Galerkin approximation for suitable approximations to $ {\mathscr H}_n$ and
${\mathscr Z}_n$ for $(n=1,2)$. We note that  
$$h(\xi,\eta)=h_0(\xi)e^{-k\eta}+kah_1(\xi)e^{-2k\eta}$$
with
$$[h, \zeta]=[{\mathscr H}_1, {\mathscr Z}_1]e^{ik(\xi-ct)}+ka [{\mathscr H}_2, {\mathscr Z}_2]
e^{2ik(\xi-ct)}.$$
Note further, $\zeta$ is obtained from the linear approximation ${\mathscr U}\zeta_\xi=-\phi_\xi$
to the kinematic surface condition ${\mathscr D}\zeta=\langle w\rangle$, which in turn yields $\zeta=-\phi/{\mathscr U}$, or
$$\zeta_1+ka\zeta_2=-{\mathscr U}^{-1}(\phi_1+ka\phi_2).$$

Following M96, S98 and SHD, and generalizing the former and the latter to the second-order Stokes waves, we arrive at
\be 
a{\mathscr P}_0=-\int_0^\infty{\mathscr U}^2({\mathscr Z}_1^{\prime 2}+k^2{\mathscr Z}_1^2)\,d\eta
-ka\int_0^\infty{\mathscr U}^2({\mathscr Z}_2^{\prime 2}+4k^2{\mathscr Z}_2^2)\,d\eta\label{5.2}
\en 
Choosing the simplest trial function for the variational integral (\ref{5.2}), namely
$${\mathscr Z}_n=ae^{-k_n\eta/b}\quad\mbox{with $k_1\equiv k$ and $k_2\equiv 2k$},$$
where $b$ is a free parameter, we obtain
$$\mathscr{P}_{n0}/k_naU_1^2=\df{b}{2}(b^{-2}+1)\left\{\df{\pi^2}{6}+\ln^2\left(\df{2k_n\gamma\eta_c}{b}\right)-2i\hat{c}_i\ln\left(\df{2k_n\gamma\eta_c}{b}\right)+\hat{c}_i^2\right\}$$
where $\hat{c}_i=c_i/U_1$, and $\gamma=0.5772$ being the Euler's number.
Proceeding as in SHD, we may cast the above expression as
$$\beta_{nc}=\pi k_n^3\eta_c^3L_{n0}^4\left\{1+\left(4-\df{\pi^2}{3}+10\hat{c}_i^2\right)\Lambda_n^2
+O(\Lambda_n^3)\right\}$$
with
$$\Lambda_n^{-1}=L_{n0}=\gamma-\ln(2k_n\eta_c).$$
Hence the energy-transfer contribution due to the initial critial layer, $\beta_c$, becomes
\be 
\beta_c&=&\beta_{1c}+ka\beta_{2c}\no\\
&=&\pi k^3\eta_c^3\left\{\left(L_{10}^4+8kaL_{20}^4\right)+\left(4-\df{\pi^3}{3}+10\hat{c}_i^2\right)
\left(L_{10}^2+8kaL_{20}^2\right)\right\}\no 
\en 

Similarly the contribution of the energy-transfer parameter due to turbulence may be evaluated from the integral (\ref{5.2}) with the extra contribution
$$\beta_T=-4k{\rm i}\int_0^\infty\lambda'\mathscr{W}\mathscr{Z}_1\mathscr{Z}'_1\,d\eta-
8k^2a{\rm i}\int_0^\infty\lambda'\mathscr{W}\mathscr{Z}_2\mathscr{Z}'_2\,d\eta$$
where $\lambda'=U_1\kappa^2$, and $\mathscr{W}=U_1\ln(\eta/\eta_c)$. Evaluating the integral it can be shown that the result may be put in the form\footnote{SHD obtained a very similar results
for monochromatic waves, namely $\beta_T=5\kappa^2L_{01}$, using a different approach.}
$$\beta_T=4\kappa^2(L_{01}+2kaL_{02}).$$

We note taht the above expression (for a monochromatic wave) is some 20\% lower than that given by SHD, (cf. equation (1.3) above).

\section{Kelvin cats-eye}

Closed streamlines, commonly known as Kelvin cats-eye, or simply cats-eye, 
occur in the neighbourhood of the critical point $\eta=\eta_c$,
where ${\mathscr U}(\eta)=0$. The stream function $\psi_b$, for
the basic flow, there
$$\psi_b\equiv\int^\eta_0{\mathscr
U}(\eta)\,d\eta\rightarrow\psi_c+\tf{1}{2}U'_c(\eta-\eta_c)^2
\qquad{\rm as}\quad\eta\rightarrow\eta_c$$
has minimum when $U'_c>0$ where
$$\psi_c\equiv\Int{0}{\eta_c}{\mathscr
U}(\eta)\,d\eta\simeq -U'_c\eta^2_c\eqno(6.1)$$ 
when ${\mathscr
U}=U_1\log(\eta/\eta_c))$. The stream function for the perturbed
flow, (\ref{5.1a}), in the neighbourhood of $\eta_c$ has the following expansion (cf.
Lighthill 1962 and Phillips 1977 \S4.3)
$$
\psi=\psi_c+\tf{1}{2}U'_c(\eta-\eta_c)^2+U'_c(\eta-\eta_c)h_c(\xi)+\phi_c(\xi),\eqno(6.2)
$$
wherein the subscript $c$ implies evaluation at $\eta=\eta_c$, and an error
factor of $1+O(ka)$ is implicit in (6.2).  

Assuming $c=c_r+{\rm i}c_i$,
\be
\phi_c=\Real\{\Phi_ce^{ik(\xi-ct)}\}
\equiv-\tf{1}{4}{\mathscr A}^2e^{kc_it}U'_c\cos
k(\xi-c_rt-x_0)\hspace*{-1.5cm}\no 
\en
$$
=\tf{1}{4}{\mathscr A}^2e^{kc_it}U_c'\left\{1-2\sin^2\left[\tf{1}{2}k(\xi-x_0)\right]\right\}\eqno(6.3)
$$
and 
$$\psi_{0}\equiv\psi_c\-\tf{1}{4}{\mathscr A}^2e^{kc_it}U'_c,\qquad\psi=\psi_c+\tf{1}{4}{\mathscr A}^2e^{kc_it}U'_c\eqno(\rm 6.4 a,b) $$
where
$$
\Phi_c\equiv\tf{1}{4}{\mathscr A}^2e^{kc_it}U'_ce^{i(\pi-kx_0-\omega_rt)}\eqno(6.5)
$$
with $\omega_r=kc_r$, then $\phi_c$ can determined from the outer solution (6.2). 
Subtracting (6.4a) from (6.4b) and substituting the result together with (6.1) and
(6.6) into (6.2), we obtain
$$
(\eta-\eta_c+h_c)^2+{\mathscr
A}^2e^{kc_it}\sin^2[\tf{1}{2}k(\xi-x_0)]=h^2_c+(2/U'_c)(\psi-\psi_0)\eqno(6.6)
$$ 
where $\psi_0\leq\psi\leq\psi_1$.

\begin{figure}
   \begin{center}
\includegraphics[width=12cm]{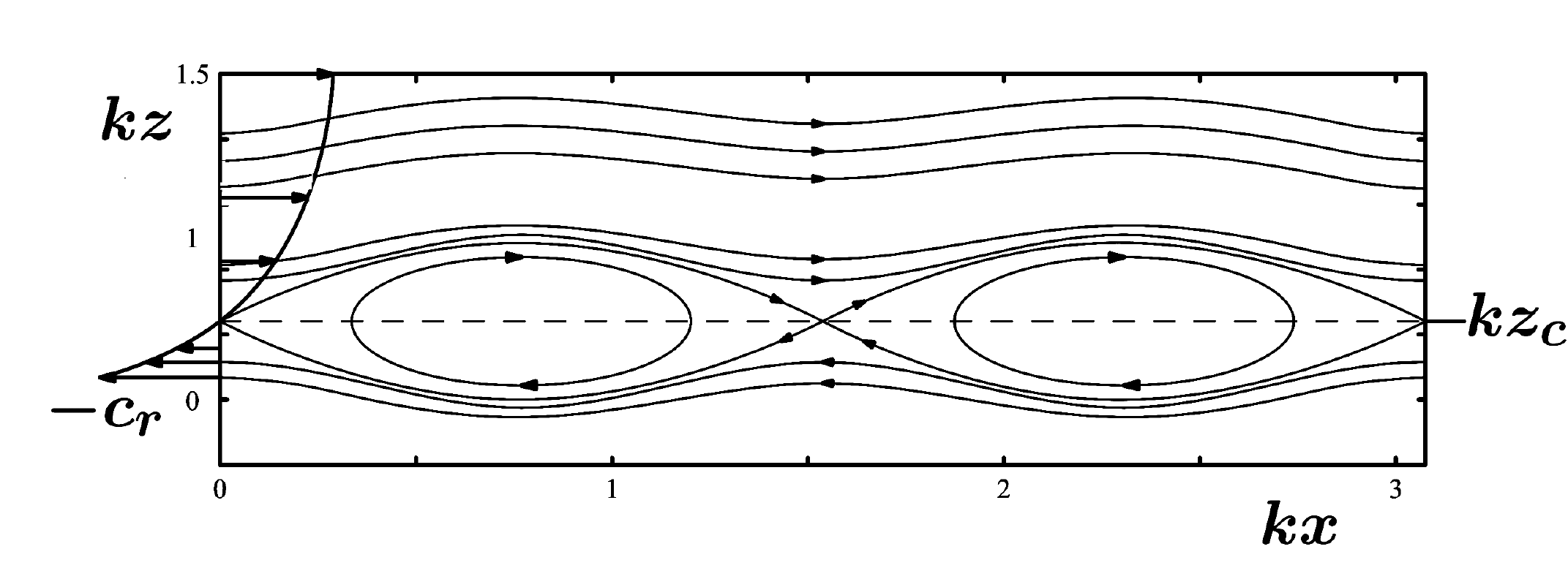}
   \end{center}
\caption{\footnotesize Formation of Kelvin's cats-eye for a logarithmic mean velocity profile over a Stokes wave.}
\end{figure}

Equation (6.6) describes a periodic sequence of nested sets
of closed streamlines, or unsteady cats-eye (on which $z=\eta+h\simeq\eta+h_c$, so that
the displacement of a particle form its ambient position $
z=\eta_c$ is $\zeta=\eta-\eta_c+h_c$) with $\psi$ as a family parameter.
If $h^2_c\ll {\mathscr A}^2e^{kc_t}$, which is usually the case, and we
assume for the sake of algebraic simplicity the centers are located at 
$k(\zeta-x_0)\,\mbox{mod}\,2\pi$ and $\eta=\eta_c$, where
$\psi=\psi_0$. Note that, the separatrix, outside of which the streamlines
are sinuous, is at $\psi=\psi_1$, and the maximum thickness of the
separatrix is $2{\mathscr A}e^{kc_t}$.  We infer from the development of
the previous section that ${\mathscr A}e^{kc_t}=O[(ka)^{1/2}\eta_c/\delta_c]$,
where $\delta_c$ is the critical-layer thickness.
These closed streamlines bifurcate from the minimum of $\psi_b$,
and the entire set may be regarded as descending from the basic
streamlines $z=\eta=\eta_c$, while the sinuous streamlines
above/below the separatrix may be regarded as descending from the
basic streamlines $z=\eta\gtrless\eta_c$ (but note that
$\psi>\psi_1$ on all sinuous streamlines).

For the purpose of demonstration, we consider an idealized case when $c_i\ll 1$ (but not exactly equal to zero). In this case the factor $e^{kc_it}\approx 1$ for a short interval 
over two cyclic wave. The result of the closed streamlines, and those above and below it are plotted in figure 2.

\section{Results and discussion}

The results of the foregoing theory warrants the following discussion.  We consider the shear flow instability where energy is transferred to surface waves by viscous mechanisms.

We point out the appliciability of the present model requires that
$$c_w<2.3(U_*/\nu_a k)^{1/3}U_*\eqno(7.1)$$
where $c_w\equiv c_r$ is the speed of the free surface waves given by
$$c_r=\left[\df{g}{k}(1+k^2a^2)+k{\cal T}\right]^{1/2}\eqno(7.2)$$
and ${\cal T}$ is the surface tension between air and water (divided by the density of the water). For all calculations reported here the crtical height is calculated from the relations
$$\eta_c=\Omega(U_1/c)^2e^{c/U_1}\qquad{\rm with}\quad\Omega=2.3\times 10^{-3}$$
The growth or decay of the Stokes surface wave initial disturbances depends on whether $c_i>0$ or $c_i<0$, respectively. Moreover, the net growth rate $\zeta=kc_i$ consists of a sum of two terms: a growth rate $\zeta_a$ due to the wind blowing over the surface of the water, and a damping rate $\zeta_w$ resulting from the viscous dissipation of the wave energy in the water.

The damping rate (for a small amplitude wave) is given by
$$\zeta_w=-2k^2\nu_w-(2k^3\nu_w c_r)^{1/2}e^{-2kd}\eqno(7.3)$$
but for deep waters (considered here) $d\gg 1$ and thus the second term in (7.3) rapidly tends to zero.

However, the growth rate due to the air, at the air-sea interface, is obtained from the solution of the Orr-Sommerfeld equation,\footnote{The mathematical detail see the appendix of Sajjadi and Drullion (2015).} see for example Benjamin (1959) by writing the solution as a sum of inviscid and viscous solutions, namely 
${\it\Phi}_n(\eta)=\phi_n(\eta)+f_n(\eta)$, for $n=1,2$. Since we have assumed $ka\ll 1$, we shall only consider the effect of the dominant harmonic (the second harmonic will have very minor effect to the overal solution), namely $n=1$, and for brevity
we drop the suffix 1. 

The inviscid solution is obtained using the method originally proposed by Miles (1962). Thus, we write
$${\it\Omega}(\eta)=(U-c)^{-1}[U'\phi-(U-c)\phi']^{-1}\phi$$
where ${\it\Omega}$ satisfies the Riccati equation
$${\it\Omega}'=k^2(U-c)^2{\it\Omega}^2-(U-c)^{-2}$$
Miles (1962) showed that
$$w=U_0'c\left[{\it\Omega}_1-\df{1}{U_1'(U_1-c)}-\int_0^{\eta_1}\df{U''\,d\eta}{U^{\prime 2}(U-c)}\right]\left[1+O(k\eta_1)^2\right]\eqno(7.4)$$
where the supscript 1 now implies evaluation at the point $\eta=\eta_1$ defined such that
$$k|\eta_c|<k\eta_1\ll 1,\qquad |U_1-c|\gg c_i$$
Upon integrating (7.4), and taking the path of integration under the singularity at $\eta=\eta_c$, we obtain
$$w_i=-\pi cU_0'\df{U''}{U_c^{\prime 3}}\left[1+O(k|\eta_c|)^2\right]$$

Following Benjamin (1959) the viscous solution satisfies
$$if^{\rm iv}+\eta f''=0$$
for which the only solution that vanishes as $\eta\uparrow\infty$ is
$$f(\eta)=\int_{\infty}^{(\eta/\delta_c)-z}\int_{\infty}^\varpi \vartheta^{1/2}H^{(1)}_{1/3}\left[\tf{2}{3}(i\vartheta)^{3/2}\right]\,d\vartheta\,d\varpi$$
where $H^{(1)}_{1/3}$  denotes the Hankel function of the first kind.  Then we may construct the following complex functions
\be 
\begin{array}{lcccl}
w(k,c)=[1+(U_0'/c)(\phi_0/\phi_0')]^{-1}, & & & &  {\mathscr F}(z)=[1+(U_0'/c)(f_0/f_0')]^{-1},\\
G(z)=f''_0/i\delta_c f'''_0, & & & {\rm and} &  H(z)=z({\mathscr F}-1)-iG\\
\end{array}\no 
\en 
The asymptotic forms of these functions, for small and large values of $z=\eta_c/\delta_c\doteqdot c/U_0'\delta_c$, is given by Miles (1960).

Thus, the growth rate, $\zeta_a$, may be determined from the viscous solution 
and can be expressed as
$$\zeta_a=\df{1}{2}sU_0'\left[\df{w_i-{\mathscr F}_i-k\delta_c(w_r-{\mathscr F}_r)H_i}{|{\mathscr F}-w|^2}\right]_{c=c_r}\eqno(7.5)$$
where $\delta_c=(\nu_a/U_0'k)^{1/3}$ and $s=1.2\times 10^{-2}$.

Using the common practice, we approximate the slope  of the mean wind velocity profile
at the surface to be
$$U_0'\thickapprox U_*^2/\nu_a$$
Furthermore, since $w_i\ll 1$ the denominator on the left-hand side of equation (7.5) may be approximated as
$$|{\mathscr F}-w|^2=({\mathscr F}_r-w_r)^2+{\mathscr F}_i^2$$
In this case, the function $w$ as given by Miles (1962), is estimated from the relation
$$w=\df{\kappa c_r}{U_*}W(R_a, A)=2.3$$
where
$$R_a=\df{\kappa U_*}{k\nu_a}\qquad{\rm and}\qquad A=\df{\kappa(u_s-c_r)}{U_*}$$
Note that, the velocity $u_s$ at the edge of the viscous sublayer, $\eta=\eta_1$, can be found from the logarithmic law of the mean velocity profile, as described by Miles (1957). Thus, the asymptotic approximation to (7.5) becomes
$$\zeta_a=\df{1}{2}sU_0'\left\{\df{w_i-(\nu_ak/2c_r)^{1/2}[(U_c'/kc_r)+2(w_r-1)]}{(w_r-1)^2+w_i^2}\right\}\eqno(7.6)$$

Having obtained the necessary formulation for the growth rates, we shall now consider the resonance between the free-surface waves in the water and the Tollmien-Schlichting waves associated with the air flow above the surface. In figure 3 we have plotted the dispersion curve $c_r/U_\infty$ as a function of $kR$ for $g=980, U_\infty=100, T=73, \nu_a=0.154$ and $\nu_w=10^{-2}$, all in c.g.s units. The Tollmien-Schlichting wave
was obtained  from the solution of the eigenvalue problem
$$c=c_2+\df{sU_0'}{k}\left[\df{1+ikH_i(z)}{{\mathscr F}'/U_0\delta_c-\partial w/\partial c}\right]\left\{(c_2+2ik\nu_w)^2-c_r^2\right\}^{-1}\eqno(7.7)$$
\begin{figure}
   \begin{center}
\includegraphics[width=12cm]{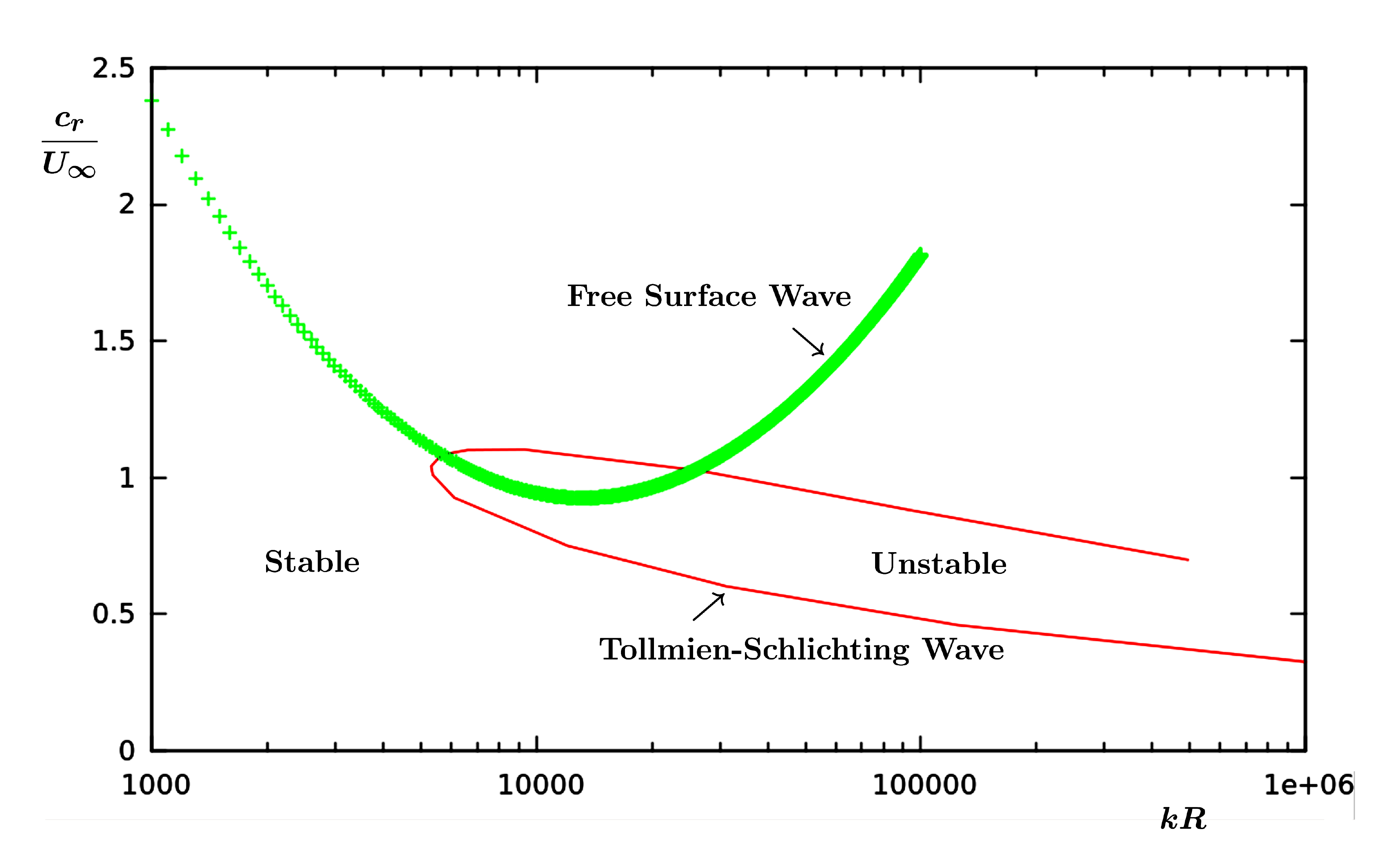}
   \end{center}
\caption{\footnotesize The neutral dispersion curve in boundary layer above the surface wave (red curve), and the neutral curve for Stokes wave on deep-water (green curve).}
\end{figure}
by equating $c_2$ to $c_r(k)$, as given by equation (7.2). The result of this graph gives a first approximation to the resonance condition between neutral ocillations associated with the mean wind velocity profile and that of deep-water waves on the free surface.

Figure 4 shows the variation of $\zeta_a$ with the wavelength $\lambda$. Also plotted is the damping rate $-\zeta_w$ for comparison. We note the peak at $\lambda\approx 3$ cm and $U_*=5$ cm/s is a close approximation to the resonant point $\lambda=2.5$ cm for $U_*=4.5$ cm/s.
\begin{figure}
   \begin{center}
\includegraphics[width=12cm]{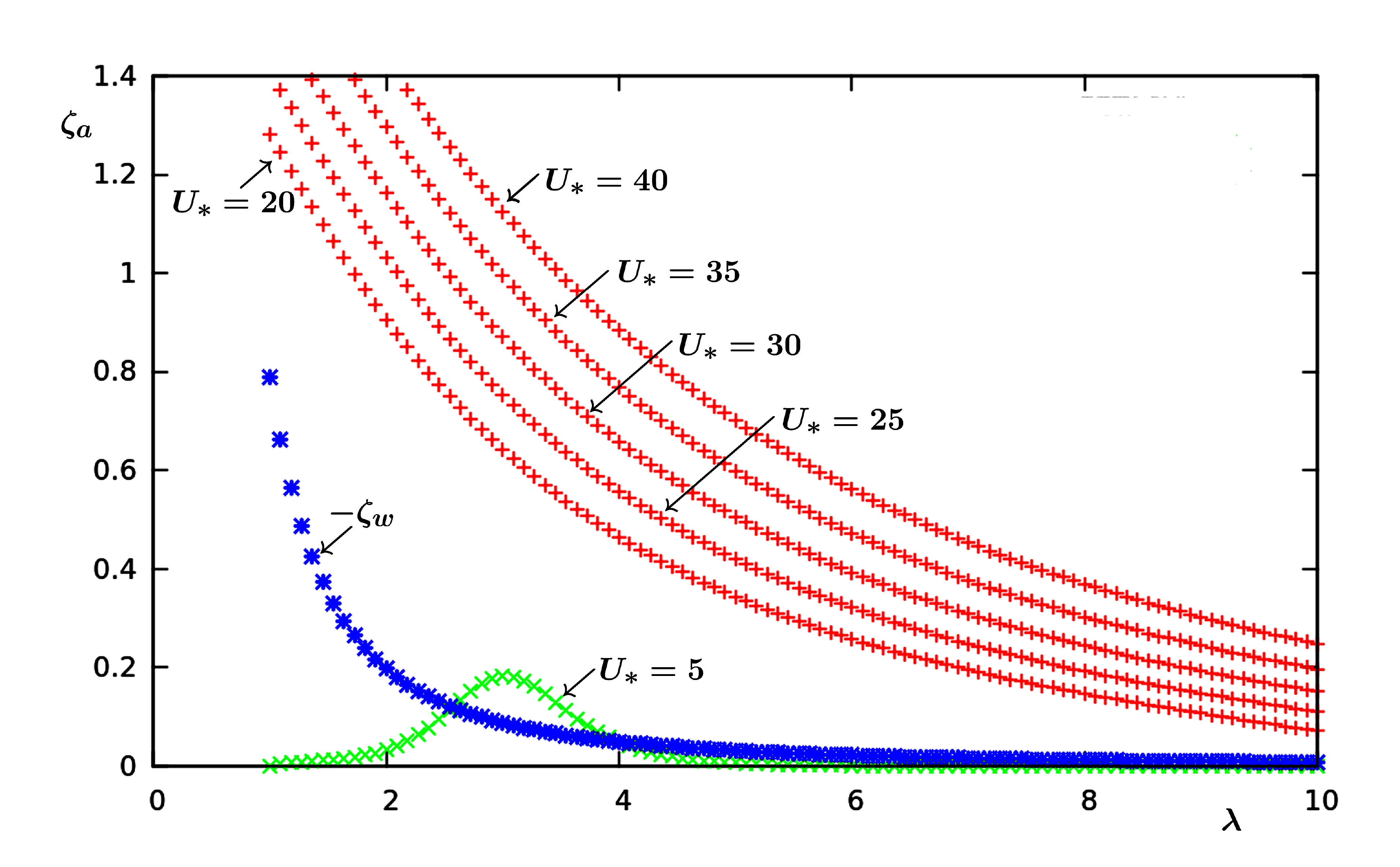}
   \end{center}
\caption{\footnotesize The growth rate for Stokes waves on deep water due to a turbulent with with logarithmic mean velocity with $U_1=5U_*$.}
\end{figure}

In figure 5 we depict the variation of $c_i$ with $\lambda$ for five values of $U_*$ in the range $20\leq U_*\leq 40$. The results show the peak value of $c_i=\zeta_a/k$ occur at lower wavelength for smaller $U_*$. From these results we deduce (i) in the absence of wind $c_i=c_r$, and (ii) the resonance appears faster for slower wind friction velocity, $U_*$, and slower for faster wind friction velocity. This condition can also be seen by plotting $c_i/U_\infty$ as a function $c_r/U_\infty$, as shown in figure 6. At very small and very large values of $c_r/U_\infty$ we observe that $\Real\{c\}\approx c_w$ which is characteristic of the Kelvin-Helmholtz instability.
\begin{figure}
   \begin{center}
\includegraphics[width=12cm]{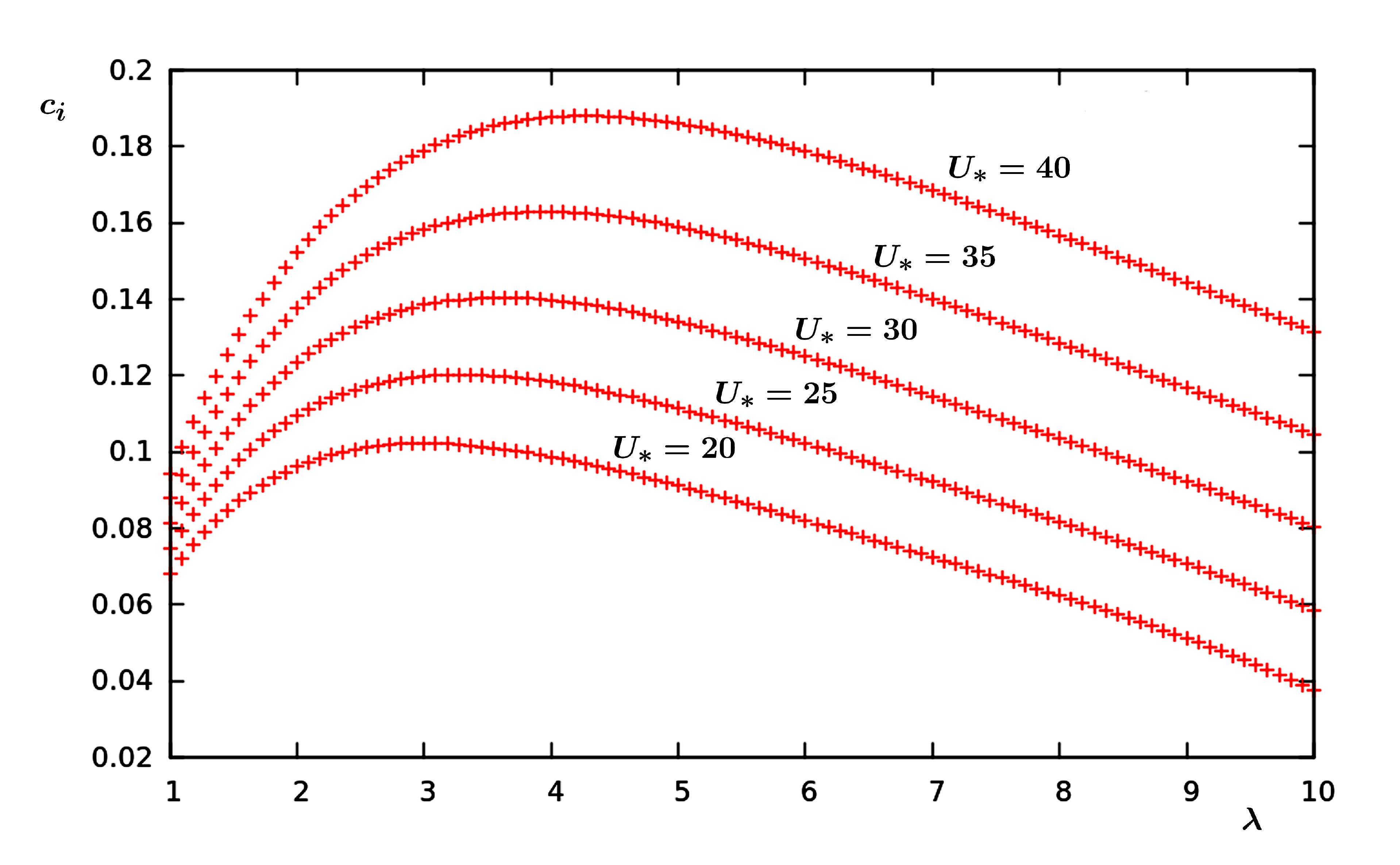}
   \end{center}
\caption{\footnotesize The variation of the complex part of the wave phase speed with the wavelength for five values of $U_*$.}
\end{figure}
\begin{figure}
   \begin{center}
\includegraphics[width=12cm]{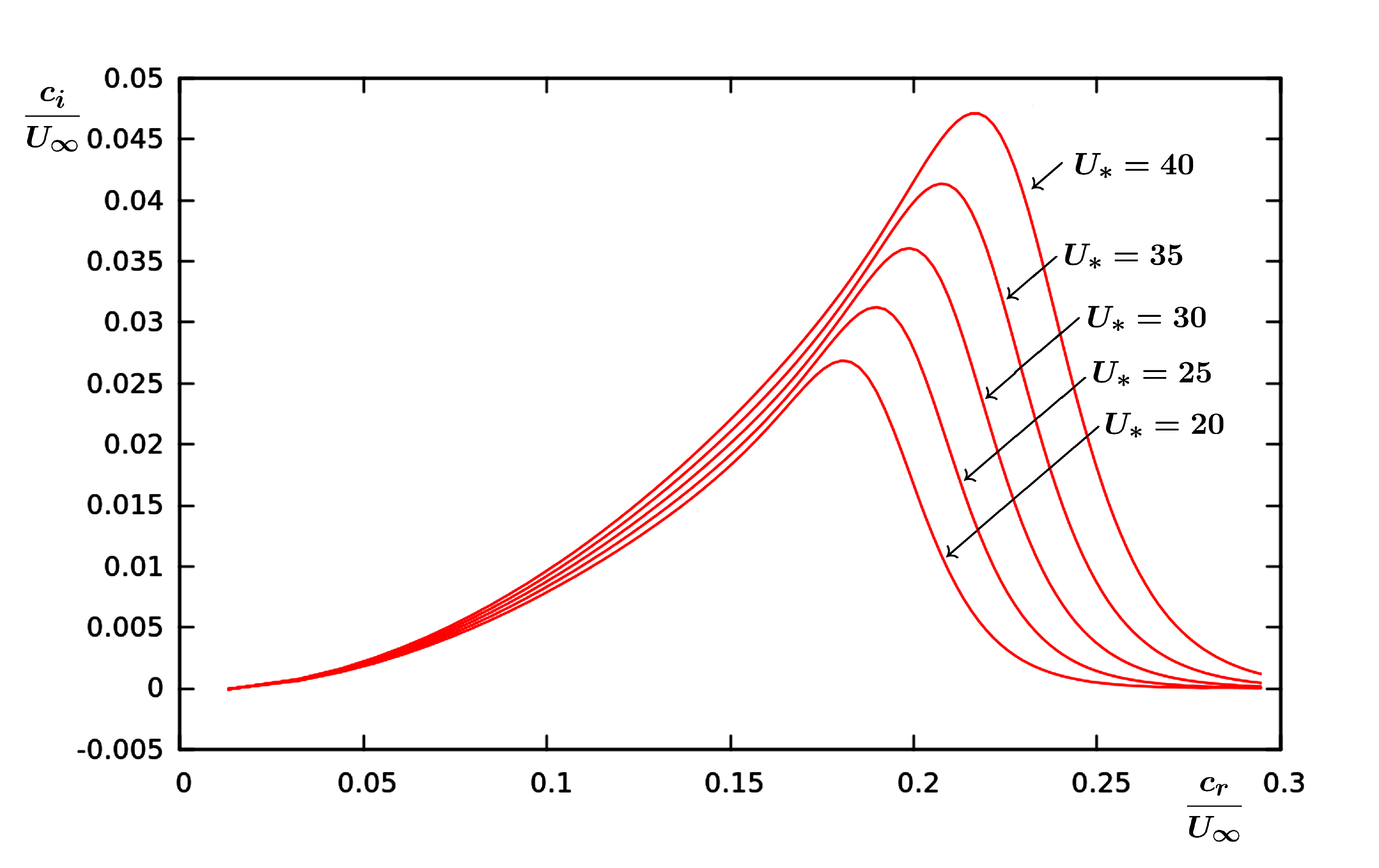}
   \end{center}
\caption{\footnotesize The variation of the normalized complex part of the phase speed with the normalized speed of a Stokes wave. The wave steepness is 0.01 and the reference wind speed is 100 cm/s.}
\end{figure}

Finally, in figure 7 we have plotted the variation of $\zeta=\zeta_a+\zeta_w$ as a function of $c_r/U_\infty$. From this figure we see that for a very slow moving wave, the growth rate is negative. This is indicative of the fact that for very small values of $c_r$ the water damping rate is dominant. As the wind, which is blowing over the wave, become stronger, which in turn increases $c_r$, the growth rate due to the air flow overcomes the viscous damping of the water. Hence, the initial surface ripples begin to grow. The maximum growth rate occurs at the resonant point, for a given value of $U_*$, and then a further increase of the wind shear (resulting to larger values of $c_r$) begings to reduce the wave growth. At this point the Tollmien-Schlichting instability will be overcome by the Kelvin-Helmholtz instability.
\begin{figure}
   \begin{center}
\includegraphics[width=12cm]{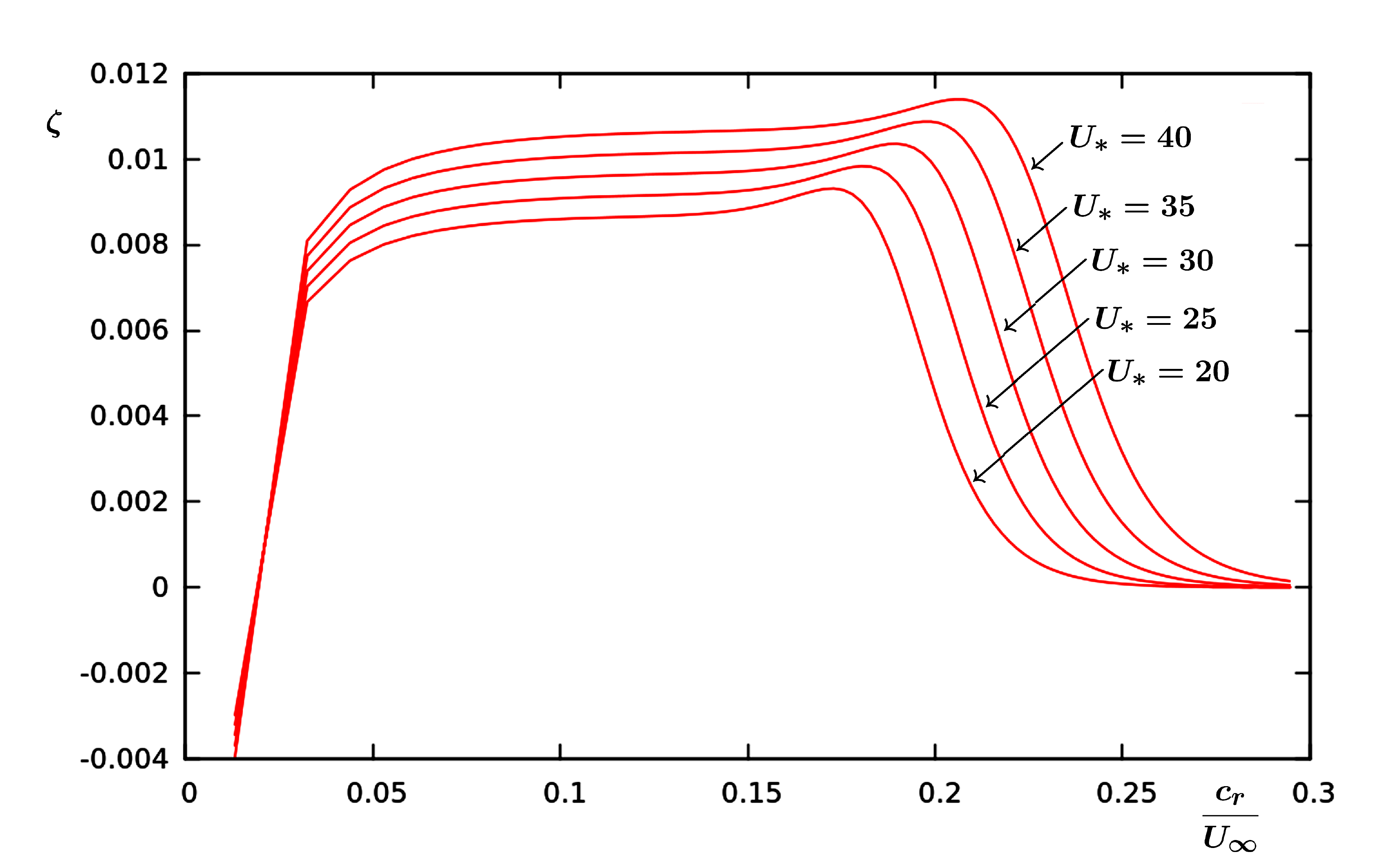}
   \end{center}
\caption{\footnotesize The variation of the net growth rate with normalized speed of Stokes wave. The wave steepness is 0.01 and the reference wind speed is 100 cm/s.}
\end{figure}   

Since the growth rate $\zeta=kc_i$ is related to the energy transfer parameter $\beta$ through
$$
\zeta=\df{1}{2}s\beta \omega_r\left(\df{U_*}{c_r}\right)^2,\qquad\omega_r\equiv kc_r
$$
and that the variation of the wave amplitude with time is given by
$$
a(t)=a_0 e^{\zeta t}
$$
Where $a_0$ is the amplitude of the wave initially (i.e. at $t=0$) we see that for the fixed value of $c_r/U_*$ as $\beta$ increases or decreases then accordingly $\zeta$ increases or decreases.   Hence, the wave amplitude accordingly grows or decays as to whether $\beta$ increases or decreases.  It should be noted that this scenario does not hold when $c_i=0$.  Thus, as a wave grows, the inertial critical layer moves up to the inner region of the flow above the wave.   Conversely, when the wave decays, the inertial critical layer moves down and for small enough values of $c_i$ it becomes confined in the inner shear layer which is very close to the wave surface. 

As the growth rate $\zeta=kc_i$ enters in the expression (6.6) we see that the position of the critical point $\eta_c$ rises or drops depending on whether $\zeta\gtrless 0$.  
Hence when $c_i\neq 0$ the cats-eye are not symmetrical above the critical point above the progressive wave.  Moreover, the thickness of the critical layer $\delta_c$ will also vary, becoming thinner when $\zeta\ll 1$ and thicker when $\zeta$ becomes larger. 

\section{Conclusions}

In our investigation here, we have considered the extension of evaluation of interfaced impedance, given originally by Miles (1957), to Stokes waves.  This task required the development of a theory of Stokes waves on a viscous fluid, which is in fact based on the original investigation of Lamb (1932).  The former which has never been previously reported in the literature shows how the interfacial impedance is affected when the surface water waves consist of multiple harmonics.  Here, the effect of the surface tension is also taken into account. 

In the theory presented here, we have assumed that the waves are growing and hence it has been assumed that the surface waves are unsteady.   In such cases, the wave phase speed is complex, when the real part represents the wave speed and the complex part is related to the growth or decay of originally formed Stokes waves by a shear flow. 

Thus, the present work extends to the previous investigation of SHD in that (i) the surface wave is unsteady and nonlinear, and (ii) the effect of the water viscosity, which affects surface stresses, is taken into account. 

For the determination of energy-transfer parameter $\beta$, we have invoked a similar turbulence closure model (the details of which will be reported in a subsequent paper), and we have shown the component of $\beta$ arising from the critical layer, namely $\beta_c$, is identical to that discovered by SHD.  However, the contribution due to turbulent shear flow $\beta_T$, is some 20\% lower than that found by SHD.  Note that $\beta_c$ essentially arises under the assumptions that (i) the critical layer is well within the inner surface layer, and (ii) the flow above the surface of water waves is inviscid (Miles 1957).  But, the interaction of turbulent shear flow with viscous water waves (which was neglected by SHD) reduces the contribution of $\beta_T$ due to viscous damping at the surface, see below.

In section 6, we have derived an expression for the closed streamlines (namely, Kelvin cats-eye, which arises in the vicinity of the critical height) when the surface wave is unsteady.  From this expression, it is clear as waves grow or decay, the cats-eye is no longer symmetrical.  We remark the symmetry only arises when the waves are steady, or more precisely, when the wave amplitude remains constant. 

Finally, we explored the energy transfer from wind to short Stokes waves through the viscous Reynolds stresses in the immediate neighborhood of the water surface.  We have conjected that the resonance between the Tollmien-Schlichting waves for a given turbulent wave profile and the free-surface Stokes waves are an additional factor that contributes to the growth of surface waves. 

\section*{Acknowledgment} This paper is dedicated to the memory of a friend and a colleague John W. Miles 1920--2007.

\end{document}